# Magnetic relaxation in bilayers of yttrium iron garnet/platinum due to the dynamic coupling at the interface

S.M. Rezende[1], R. Rodríguez-Suárez[2], M. M. Soares[1], L. H. Vilela-Leão[1], and A. Azevedo[1]

[1]Departamento de Física, Universidade Federal de Pernambuco, 50670-901, Recife, PE, Brasil.
[2]Facultad de Física, Pontificia Universidad Católica de Chile, Casilla 306, Santiago, Chile.

We show that in ferromagnetic (FM)/normal metal (NM) bilayers the dynamic coupling at the interface transfers an additional magnetic relaxation from the heavily damped motion of the conduction electron spins in the NM layer to the FM spins. While the FM relaxation rates due to two-magnon scattering and spin pumping decrease rapidly with increasing FM film thickness, the damping due to the dynamic coupling does not depend on the FM film thickness. The proposed mechanism explains the very large broadening of ferromagnetic resonance lines in thick films of yttrium iron garnet after deposition of a Pt layer.

One of the fundamental properties of a magnetic system is the manner by which its magnetization relaxes towards equilibrium. This is governed by the spin interactions and the structure of the magnetic system and its detailed understanding is important from the point of view of basic physics and for technological applications. For several decades the magnetic relaxation has been investigated experimentally in bulk and thin film materials mainly by measuring the linewidth of the ferromagnetic resonance (FMR) at microwave frequencies. In bulk magnetic insulators the relaxation occurs through intrinsic mechanisms involving magnon-magnon and magnon-phonon processes as well as extrinsic mechanisms such as scattering by impurities.[1,2] In bulk metallic materials the relaxation is dominated by processes involving the conduction electrons.[3] In very thin films and multilayers new physical relaxation processes have been discovered in the last fifteen years, the most important ones being two-magnon scattering from the irregularities at the surfaces or interfaces[4,5] and the spin pumping mechanism.[6,7] These processes contribute with additional relaxation rates that increase as the magnetic film thickness decreases and thus become very important in ultra-thin films.[5,8]

In recent years structures made of bilayers of ferromagnetic metal (FM) / normal metal (NM) films have been attracting considerable interest due to the discoveries of the spin Hall effect[9,10] and the inverse spin Hall effect (ISHE).[11,12] In a FM/NM bilayer undergoing ferromagnetic resonance (FMR) it has been found[11-14] that the precessing spins in the FM inject spins into the adjacent NM layer generating a spin-pumping dc voltage by means of the ISHE opening immense possibilities in the field of spintronics.[15] A very important recent development in this field was the demonstration that the ferrimagnetic insulator yttrium iron garnet (YIG) can be used in FM/NM structures to convert charge current into spin current and vice-versa.[16] Due to its small magnetic damping, YIG films can be used to transport spin information over much larger distances than in FM metals so that YIG/Pt structures have attracted increasing scientific attention.[17-26] However it has been observed that the deposition of a Pt layer on thick YIG films produces an unusually large broadening of the microwave absorption lines,[17,18] which is quite surprising because one expects the spin pumping mechanism to be effective only at ultra-thin films.

In this paper we show that when a NM layer is deposited on a FM film, in addition to the spin pumping process there is another mechanism for magnetic relaxation which is effective in thick FM films. The mechanism relies on the transferred relaxation due to the dynamic coupling of the precessing magnetization in the FM with the heavily damped precession of the conduction electron spins in the NM layer. This process is effective in FM metallic or insulating films and is independent of the spin-pumping mechanism, although both originate in the spin coupling at the interface. While the spin-pumping mechanism is due to the flow of angular momentum out of the FM layer into the NM layer and relaxes the longitudinal component of the magnetization, the new mechanism relaxes the transverse components of the magnetization. We show that the relaxation due to dynamic coupling at the interface explains the observed broadening of the FMR lines in thick YIG films with deposition of a Pt layer.

We consider a bilayer of a ferromagnetic material with a nonmagnetic metal in which the precessing magnetic moments of the FM layer interact with the heavily damped spins of the conduction-electron spins in the NM layer through the dynamic exchange coupling at the interface. In order to treat the coupled mode problem we follow the macroscopic approach of Ref. [16] and consider that at the interface sites $i$ the spins $\vec{s}_i$ of the conduction electrons in the NM layer interact with the spins $\vec{S}_i$ in the FM side through the s-d exchange interaction, $H_{sd} = -J_{sd} \sum_i \vec{S}_i \cdot \vec{s}_i$, where $J_{sd}$ is the exchange coupling constant. Writing the relation between the magnetization and the spins as $\vec{M}(\vec{r}) = g\mu_B \sum_i \vec{S}_i \delta(\vec{r}-\vec{r}_i)$, where g is the Landé factor and $\mu_B$ is the Bohr magneton, the summation on the interface sites $i$ can be written as a surface integral along the interface and the coupling between the magnetization $\vec{M}(\vec{r},t)$ in the FM side and the magnetization $\vec{m}_N(\vec{r},t)$ of the conduction electrons in the NM side can be written as,

$$H_{sd} = -(J_{ex}/A) \int dx\, dz \int dy\, \vec{M}(\vec{r},t) a_{eff} \delta(y) \cdot \vec{m}_N(\vec{r},t), \quad (1)$$

where $y$ is the direction perpendicular to the interface plane $x$-$z$ with area $A$ at $y = 0$, $J_{ex} = J_{sd} S/(\hbar \gamma_e M)$ is the dimensionless exchange coupling constant, $S$ is an effective block spin per unit cell and $M$ is the magnetization of the FM, $\gamma_e$ is the gyromagnetic ratio of the conduction electrons in the NM, $a_{eff} = v_e/a_s^2$ is the effective interaction range, $v_e$ is the volume per conduction electron, and $a_s$ is the lattice constant of the localized spins at the interface on the FM side. In order to make the interface coupling tractable we follow Ref. [16] and consider that the magnetizations do not vary along the interface plane and $\vec{m}_N(\vec{r},t) = \vec{m}_N(y,t)$, so that from Eq. (1) we obtain,

$$H_{sd} = -\zeta J_{ex} \vec{M}(t) a_{eff} \cdot \vec{m}_N(y=0,t), \quad (2)$$

where $\zeta$ is a factor that accounts for the surface integral: $\zeta = 1$ corresponds to an atomically flat surface and uniform magnetizations at the interface plane; $\zeta < 1$ accounts for irregularities at the interface, such as roughness, or to a spatially varying $\vec{M}(\vec{r},t)$ as in a spin wave. Equation (2) represents the interface energy per unit area and since the magnetization is distributed over the whole FM volume, the energy per unit volume on the FM side is $E_v^{FM} = -\zeta J_{ex} \vec{M}(t) a_{eff} \cdot \vec{m}_N(0,t)/d_{FM}$, where $d_{FM}$ is the FM layer thickness. One can write the effective field acting on the FM magnetization due to the interface exchange as,

$$\vec{H}_E^{FM} = -\frac{\partial E_v^{FM}}{\partial \vec{M}} = \zeta J_{ex} a_{eff} \vec{m}_N(0,t)/d_{FM}. \quad (3)$$

$$\frac{d\vec{M}}{dt} = -\gamma \vec{M} \times \vec{H} - \gamma \zeta J_{ex}[\vec{M} \times \vec{m}_N(0)]\frac{a_{eff}}{d_{FM}} - \eta_{FM} \vec{M}, \quad (4)$$

where $\gamma$ is the gyromagnetic ratio ($g\mu_B/\hbar = 2\pi \times 2.8$ GHz/kOe for YIG) and $\eta_{FM}$ is the relaxation rate of the FM magnetization in the absence of coupling at the interface. The equation of motion for the magnetization $\vec{m}_N(y,t)$ of the conduction electrons must take into account the spin diffusion into the NM side and the exchange coupling at the interface. The magnetization in the NM can be written[16] as $\vec{m}_N(y) = \vec{m}_0 a_{eff} \delta(y) + \delta\vec{m}_N(y)$ where $\vec{m}_0 = \chi_N \zeta J_{ex} \vec{M}$ is the equilibrium magnetization, $\chi_N$ is the paramagnetic susceptibility of the conduction electrons and $\delta\vec{m}_N(y,t)$ represents the spin accumulation. The equation of motion then becomes,[16]

$$\frac{\partial \vec{m}_N}{\partial t} = -\gamma_e \vec{m}_N \times \vec{H} - \gamma_e \zeta J_{ex}(\vec{m}_N \times \vec{M}) a_{eff} \delta(y) \\ - \eta_{sf} \vec{m}_N + D_N \nabla^2 \delta\vec{m}_N, \quad (5)$$

where $\eta_{sf}$ is the relaxation rate of the spin accumulation, related to the electron spin-flip time $\tau_{sf}$ by $\eta_{sf} = 1/\tau_{sf}$, and $D_N$ is the spin diffusion constant. We consider that the static field is in a direction parallel to the interface plane, designated the z-direction of a coordinate system that has the y-direction perpendicular to the interface and write the two magnetizations as $\vec{M} = \hat{x} m_x + \hat{y} m_y + \hat{z} M_z$ and $\delta\vec{m}_N = \hat{x}\delta m_N^x + \hat{y}\delta m_N^y + \hat{z}\delta m_N^z$. Equations (4) and (5) describing the coupled motion of the magnetizations can be solved for all magnetization components. This has been done in Ref. [16] for the longitudinal spin accumulation $\delta m_N^z(y,t)$ from which one can calculate the spin current density in the NM using $J_s^z = -(e/\mu_B) D_N \nabla_y \delta m_N^z$, which leads to, in units of angular momentum/(area·time),

$$J_s^z(0) = \frac{\omega \hbar^2 \chi_N D_N}{2\mu_B^2 \lambda_N (1+\Gamma^2)} \left(\frac{m_x m_y}{M_z^2}\right), \quad (6)$$

where $\lambda_N = (D_N \tau_{sf})^{1/2}$ is the spin diffusion length, $\Gamma$ is the parameter defined in Ref. [16], with $\zeta = 1$, $\Gamma = \hbar \lambda_N/(S J_{sd} \tau_{sf} a_{eff})$ and we have assumed circular precession. Comparison of Eq. (6) with the well known expression for the current density in terms of the transverse components of the magnetization[6,7] leads to a convenient relation between the exchange coupling parameter and the spin-mixing conductance $g^{\uparrow\downarrow}$,

$$g^{\uparrow\downarrow} = \frac{2\pi \hbar \chi_N D_N}{\mu_B^2 \lambda_N (1+\Gamma^2)}. \quad (7)$$

Notice that this relation differs by a factor of 2 from the one in Ref. [16] because we have considered $m^+ m^- = m_x^2 + m_y^2 \approx 2 m_x m_y$. The spin current in Eq. (6) represents a flow of spin angular momentum out of the FM layer resulting in the relaxation of the FM magnetization through the spin pumping mechanism giving rise to a FMR linewidth (half width at half maximum) given by[6,7]

$$\Delta H_{SP} = \frac{\hbar \omega g^{\uparrow\downarrow}}{4\pi M d_{FM}}. \quad (8)$$

where $4\pi M$ is the saturation magnetization. Note that the spin pumping process relaxes the $M_z$ component of the magnetization so that it produces a relaxation time of the type $T_1$ in the Bloch-Bloembergen formulation[1,2] of the Landau-Lifshitz equation. In the derivation of Eq. (6) one neglects [16] the reaction of the spin accumulation $\delta \vec{m}_N$ on the FM magnetization $\vec{M}$. The full solution of Eqs. (4) and (5) for the coupled transverse components of the magnetizations reveals another independent contribution to the FM relaxation of the type $T_2$ arising in the dynamic coupling between the spins at the interface. Using for the transverse variables $\delta m_N^+ = \delta m_N^x + i \delta m_N^y$ and $m^+ = m_x + i m_y$ we can solve the equation for the spatial dependence of the transverse spin accumulation $\delta m_N^+(y,t)$ obtained from Eq. (5). Considering for simplicity a NM layer thickness much larger than the spin diffusion length we find $\delta m_N^+(y,t) = \delta m_N^+(0,t) \exp(-y/\lambda_N)$ so that from Eq. (5) we obtain an equation relating $m^+(t)$ and $\delta m_N^+(0,t)$. This equation together with the one obtained from Eq. (4) form a set of two equations for the transverse variables. Considering the time dependence $\exp(i\omega t)$ one obtains the coupled equations,

$$[(\omega - \omega_{FM}) - i\eta_{FM}]m^+ = -\gamma M_z (\zeta J_{ex} a_{eff}/d_{FM}) \delta m_N^+(0), \quad (9)$$

$$[(\omega - \omega_H)/2 - \lambda_{ex}\eta_{sf}(1-i)]\delta m_N^+(0) = \\ -\lambda_{ex}[(\omega-\omega_H)/\gamma M_z]\chi_N \eta_{sf} m^+, \quad (10)$$

where $\omega_H = \gamma_e H$ is the conduction electron spin resonance frequency, $\omega_{FM} = \gamma\sqrt{H(H+4\pi M)}$ is the FMR frequency and $\lambda_{ex} = \zeta/\Gamma$ is a dimensionless coupling parameter. From Eqs. (9) and (10) one obtains,

$$[(\omega-\omega_{FM})-i\eta_{FM}][(\omega-\omega_{FM})+\Delta\omega_{FM}-2\eta_{sf}(i+\lambda_{ex})] \\ -2[(\omega-\omega_{FM})+\Delta\omega_{FM}]C = 0 \quad (11)$$

where $\Delta\omega_{FM} = \omega_{FM} - \omega_H$ and $C = m_0 a_{eff}\eta_{sf}\lambda_{ex}/(M d_{FM})$. Eq. (11) leads to a quadratic equation with complex coefficients, the roots of which are the two complex eigenmode frequencies,

$$\omega_1 \approx \omega_H + i 2\eta_{sf}, \quad (12)$$

$$\omega_2 \approx \omega_{FM} + i[\eta_{FM} + \lambda_{ex}^4 \eta_{sf} + 4\lambda_{ex}^4 \eta_{sf}(\frac{m_0}{M})(\frac{a_{eff}}{d_{FM}})]. \quad (13)$$

The real and imaginary parts of Eqs. (12) and (13) correspond respectively to the eigenmode oscillation frequencies and relaxation rates. Clearly $\omega_1$ is associated with the motion dominated by the conduction electron spins in the NM layer, whereas $\omega_2$ is associated with the spin precession in the FM layer. Since $\omega_H \sim 10^{10}$ s$^{-1}$ and $\eta_{sf} \sim 10^{12}$ s$^{-1}$ the motion of the spins in Pt is heavily overdamped. The important result revealed in Eq. (13) is that the relaxation rate of the FM layer has, in addition to the intrinsic term $\eta_{FM}$, two contributions proportional to the fourth power of the exchange coupling parameter and to the conduction electron spin relaxation $\eta_{sf}$. Note also that since $m_0/M < 1$ and $a_{eff}/d_{FM} \ll 1$, for YIG films with μm thickness, the third term in the imaginary part of Eq. (13) is negligible compared to the second. As will be shown in the following, the damping transferred from the conduction electron spins in the NM layer to the FM magnetization represented by the second term in the imaginary part of Eq. (13) explains the large broadening of the FMR lines observed in YIG/Pt.

The experiments were carried out at room temperature with several samples consisting of a single-crystal YIG film with thickness in the range 8 to 28 μm, grown by liquid-phase epitaxy (LPE) on 0.5 mm thick (111) gadolinium gallium garnet substrates and cut in an approximate square shape with lateral dimensions of 2 to 4 mm. Measurements were made with the bare YIG films and with a Pt layer with thickness 6 nm or 20 nm deposited by magnetron sputtering. The samples are mounted at the tip of a plastic rod to allow measurements as a function of the angle and placed in the middle of a shorted rectangular waveguide between the poles of an electromagnet. The static magnetic field with intensity $H$ and the microwave magnetic field are kept perpendicular to each other. Field sweep measurements were made to obtain spectra of the microwave absorption derivative at several values of the angle $\theta_H$ between the field $\vec{H}$ the film plane to investigate the out-of-plane angle dependence of the spectra.

All samples investigated exhibited similar behavior. The FMR linewidths in the bare YIG film are less than 1 Oe with the applied field parallel or perpendicular to the film plane. After deposition of a 6 nm or 20 nm Pt layer the linewidths increase to several Oe. We present here only some representative results obtained with microwave frequency 9.4 GHz and input power less than 10 mW. Fig. 1 shows the field derivative d$P$/d$H$ of the microwave absorption spectra for a 28 μm thick YIG film for two directions of the field, parallel and perpendicular to the film plane. The spectra in Figs. 1(a) and 1(b) obtained before deposition of the Pt layer allow a clear identification of the absorption lines. They correspond to standing spin-wave modes that have quantized in-plane wave numbers due to the boundary conditions at the edges of the film.[17,18,27] The strongest line corresponds to the FMR mode that has frequency given by $\omega_0 \approx \gamma H^{1/2}(H+4\pi M)^{1/2}$. In Fig. 1(a) with the field in the plane, the lines to the left of the FMR mode correspond to hybridized surface modes whereas those to the right are volume magnetostatic modes.[17,18] As shown in Figs. 1(c) and (d) the deposition of a 6 nm thick Pt layer on the YIG film produces a pronounced broadening of the lines for all modes. They have nearly the same peak-to-peak linewidth $\Delta H_{pp}$ of about 6 Oe, which is nearly ten times larger than in bare YIG. Similar results were obtained with YIG film thickness 8 and 15 μm. In both samples the linewidths in YIG/Pt have nearly the same value for $\theta_H = 0$ and 90°, which are respectively 2.5 and 3.0 Oe larger than in the corresponding bare YIG film.

The first mechanism one would consider to explain the increased damping in YIG caused by deposition of Pt is the spin pumping process.[6,7,23] Using for the effective spin-mixing conductance the value $g_{\uparrow\downarrow} \approx 1.5\times10^{13}$ cm$^{-2}$ determined in Ref. [18] from measurements of $V_{SP}$ using the spin-Hall angle $\gamma_H = 0.08$ and spin-diffusion length $\lambda_N = 3.7$ nm,[14,18] and $4\pi M = 1.76$ kG for YIG, we obtain with Eq. (8) for a 28 µm thick YIG/Pt bilayer a linewidth of $\Delta H_{SP} = 4\times10^{-4}$ Oe. This is definitely too small compared to the observed line broadening ruling out the spin pumping as the source of the additional relaxation. Another possible origin of the line broadening with Pt deposition is the enhancement in the interface two-magnon scattering relaxation similar to the one observed when a FM layer is in contact with an antiferromagnetic material.[5] However, the two-magnon linewidth[4] varies with $1/d_{FM}^2$ and like the spin pumping it is important only in films with thickness of a few nanometers. Moreover, with the field normal to the plane the FMR frequency is at the bottom of the spin-wave manifold and there are relatively few degenerate states into which the FMR mode can decay into, so that the relaxation due to two-magnon scattering should be small as demonstrated theoretically and observed experimentally.[29,30] As it is clear in Fig. 1, the linewidths in YIG/Pt at $\theta_H = 0$ and 90° are very similar, ruling out the two-magnon scattering as the mechanism responsible for the line broadening.

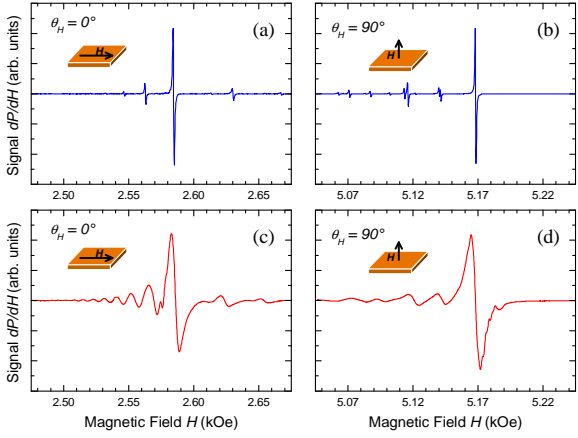

**Figure 1**. Field scan microwave field derivative absorption spectra at a frequency 9.4 GHz of a 28 µm thick YIG film with lateral dimensions 2 x 3 mm with the magnetic field applied parallel or perpendicular to the film plane as indicated. In (a) and (b) the YIG film is bare while in (c) and (d) it is covered with a 6 nm Pt layer.

We will show now that the transferred relaxation mechanism proposed here accounts comfortably for the observed line broadening in YIG/Pt. For FM films with thickness in the µm range the additional damping in Eq. (13) is dominated by the second term in the imaginary part so one can write the contribution of the transferred relaxation to the FMR linewidth as,

$$\Delta H_{trans} = \lambda_{ex}^4 \eta_{sf} / \gamma, \quad (14)$$

where the $\lambda_{ex} = \zeta/\Gamma \ll 1$ exchange parameter is related to the spin mixing conductance by,

$$\lambda_{ex}^2 = g^{\uparrow\downarrow} \frac{\zeta^2 \mu_B^2 \lambda_N}{2\pi\hbar \chi_N D_N}. \quad (15)$$

Atomic force microscopy analysis has revealed that the surfaces of the LPE grown YIG films are flat in the atomic level, so we consider $\zeta = 1$. Using the spin mixing conductance[18] for YIG/Pt, $g_{\uparrow\downarrow} = 1.5\times10^{13}$ cm$^{-2}$, and the following parameters for Pt, $\lambda_N = 3.7$ nm (Ref.14), $\eta_{sf} = 10^{12}$ s$^{-1}$ (Ref.16), $D_N = \lambda_N^2 \eta_{sf} = 0.1$ cm$^2$s$^{-1}$, $\chi_N = 2.1\times10^{-5}$ (paramagnetic susceptibility of bulk Pt, Ref.31), we obtain $\lambda_{ex}^2 = 0.024$. This gives for the theoretical additional FMR linewidth due to the transferred relaxation a value $\Delta H_{trans} \approx 32$ Oe. This value is six to ten times larger than the measured ones, revealing that the dynamic interaction at the interface indeed transfers a considerable damping from the overdamped motion of the conduction electron spins in Pt to the magnetization in YIG. The discrepancy between theory and experiment can be attributed to several factors. First we note that there is a large discrepancy in the parameters for Pt reported by different authors.[32] Second, and perhaps more important, we have used in the calculation of $\lambda_{ex}$ the value of the paramagnetic susceptibility for bulk Pt, which is the only one available in the literature.[31] However, there are experimental evidences that when Pt is deposited on a FM metal, such as Ni and Py, the atomic layers close to interface become spin polarized due to the proximity effect.[33,34] This effect most certainly occurs also in YIG/Pt, producing an enhancement of the Pt susceptibility near the interface, which decreases the value of $\lambda_{ex}$ and consequently lowers the theoretical prediction for $\Delta H_{trans}$. Note that an enhancement by a factor of 3 in the Pt susceptibility, which is well within the increase in the magnetic moment near the interface measured is Refs. [33,34], is sufficient to produce a nice agreement between theory and experiments.

In conclusion, we have demonstrated that in FM/NM bilayers, the coupling of the overdamped motion of the conduction electron spins in the NM layer to the spins in the FM layer gives rise to a transferred damping mechanism. This additional damping has an origin that is independent of the well known spin-pumping mechanism.[6,7] Both relaxation processes originate in the interface coupling but they are due to quite different mechanisms. The conventional spin-pumping is due to the flow of angular momentum out of the FM layer into the NM layer and relaxes the longitudinal component of the magnetization. It does not depend on the conduction electron spin-flip relaxation rate $\eta_{sf}$ and varies inversely with the thickness of the FM layer. On the other hand, the mechanism proposed here relaxes the transverse components of the magnetization, resulting in a damping rate proportional to $\eta_{sf}$ and independent of the thickness of the FM layer. The existence of this novel relaxation mechanism is clearly observed in the broadening of the ferromagnetic resonance lines in relatively thick YIG films after deposition of a Pt layer.

This work was supported in Brazil by the agencies CNPq, CAPES, FINEP and FACEPE and in Chile by the Millennium Science Nucleus "Basic and Applied Magnetism" No. P10-061-F.